\documentclass[12pt]{article}
\usepackage[margin=2cm,nohead,a4paper]{geometry}

\usepackage{graphicx}
\usepackage{latexsym}
\usepackage{amsfonts,amsmath,amssymb}
\usepackage{url}
\usepackage{fancyref}
\usepackage{natbib}
\usepackage[noblocks]{authblk}

\begin{document}

\title{Modeling Light Curves for Improved Classification}

\author[1]{Julian Faraway}
\author[2]{Ashish Mahabal}
\author[3]{Jiayang Sun}
\author[4]{Xiaofeng Wang}
\author[5]{Yi (Grace) Wang}
\author[6]{Lingsong Zhang}

\affil[1]{University of Bath}
\affil[2]{California Institute of Technology}
\affil[3]{Case Western Reserve University}
\affil[4]{Cleveland Clinic Lerner Research Institute}
\affil[5]{SAMSI and Duke University}
\affil[6]{Purdue University}

\maketitle

\begin{abstract}
  Many synoptic surveys are observing large parts of the sky multiple
  times.  The resulting lightcurves provide a wonderful window to the
  dynamic nature of the universe.  However, there are many significant
  challenges in analyzing these lightcurves.  We describe a
  modeling-based approach using Gaussian process regression for
  generating critical measures for the classification of such
  lightcurves. This method has key advantages over other popular
  nonparametric regression methods in its ability to deal with
  censoring, a mixture of sparsely and densely sampled curves, the
  presence of annual gaps caused by objects not being visible
  throughout the year from a given position on Earth and known but
  variable measurement errors. We demonstrate that our approach
  performs better by comparing it with past methods based on summary
  statistics.  Finally, we provide future directions for use in
  sky-surveys that are getting even bigger by the day.

\end{abstract}

\bibliographystyle{abbrvnat}

\section{Introduction}
\label{sec:introduction}

In the last few decades we have seen advances in imaging technology,
and the storage, transfer and processing of data.  As a result,
astronomy has moved from taking static, sporadic snapshots of the sky
to obtaining high-cadence, deep and large images, almost akin to
making digital movies of the sky. This, in turn, has resulted in opening up
the field of studying the dynamic nature of the universe, in
particular, the cataloging of different types of objects, both within
our Galaxy, and all the way to the early
Universe. Cataloging goes well beyond stamp-collecting, since it
reveals the time scales over which various phenomena occur, directly
relating to the physical processes behind the brightness changes in
astronomical objects, and allowing us to connect the different
families of objects in various ways. A bonus is also the ability to
look for connections missing so far, as well as fringe members of
different classes.

Much of characterizion or classification for cataloging is done, or at least begins,
through the study of variability of objects. Most astronomical
objects, be they stars or planets or galaxies, or any of their
subclasses, vary in brightness either intrinsically through some
physical process such as explosion, merging or infall of matter, or
through an extrinsic process such as eclipse or rotation. For a small
fraction of objects the variation can happen over a fraction of a
second to hundreds of days depending on the phenomenon. For a majority
of objects, the changes are much slower and smaller as the objects
evolve through the proverbial astronomical time-scales. We can observe
large parts of the sky multiple times at different wavelengths, yet
these observations are far from continuous, all-sky, or
panchromatic. For each part of the sky, and in particular for each
object in the part of the sky we image, we get a time-series of
flux. While all objects vary to an extent, for a vast majority of
objects, the variations are non-discernible during the rather sparse
sequence (tens to hundreds of epochs) of short exposures (less than a
minute) that we have, and over the time-scales over which observations
occur (few years). That is precisely the reason, for instance, that
when we glance at the night sky we do not find stars suddenly changing
their brightness.

This leads to most astronomical objects seeming {\em
  non-variable}. When we can discern the variability, e.g. a periodic
variation, or a stochastic variation, or even a single sudden jump in
brightness, the object could then be called a {\em variable}. This
functional definition would of course change based on many factors,
such as, total interval of observation, type of phenomena involved
etc. An extreme case of a variable object is a \textit{transient} - the
brightness of which varies by several standard deviations in a much
shorter time, of the order of seconds to minutes. It is the study of
these types of objects that has really become possible due to
high-cadence wide-field surveys.

In order to understand and classify transients, it is important to
understand variability at all levels, including mostly non-variable
astronomical sources.  Past attempts have included analyses for denser
lightcurves from Kepler, as in \citet{Blomme2010}, \citet{Ciardi2011}
or using brighter objects as in \citet{Richards2011} and general
frameworks based on such approaches as in
\citet{mahabal2011discovery}, \citet{Djorgovski2012} as well as
\citet{Bloom2012}, \citet{Richards2012}, \citet{Blocker2013},
\citet{Graham2014}. It is important to design measures that can
isolate specific classes but are also derivable based on the available
cadence of observations. Our aim is to present new measures based on
object lightcurves which help in better discriminating between
variables and non-variables, and among the different transient types.
See \citet{borne2009scientific}) for an application to larger datasets and
\citet{peng2012selecting} for use on specific classes.

Here we use data from the Catalina Real-time Transient Survey (CRTS)
(\citet{Drake2009}). CRTS is based on the Catalina Sky Survey (CSS)
which has been designed to look for near-Earth asteroids. One way to
look for asteroids is by looking for motion of the asteroids with the
backdrop of mostly non-moving stars in the night sky. The cadence used
for this is four images taken 10-minutes apart. Thus, the CRTS
lightcurves have 4 points obtained within 30 minutes. The next such
set could be the next night, the next week or even a month later. The
sparse and non-uniform nature of the lightcurves presents
classification challenges and also allows development of new
statistical techniques.

\begin{figure}[ht]
  \centering
  \includegraphics*[height=3in]{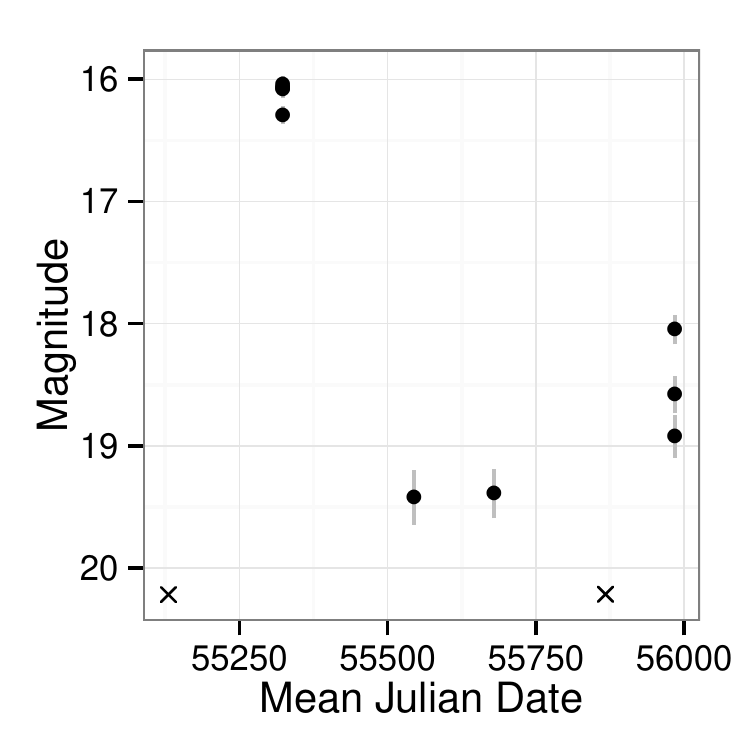}
  \caption{Fragment of a lightcurve. The solid dots represented recorded observations and are plotted with error bars. The
    two crosses represent times when an observation was attempted but only an upper limit on the magnitude can be specified.
  \label{fig:eglc}}
\end{figure}

A small section of about one year of light
curve observation is shown in Figure~\ref{fig:eglc}. The magnitude is
the negative logarithm of flux so in keeping with standard practice,
we plot the magnitudes on a reversed scale because smaller magnitudes
represent brighter objects. We see two groups where multiple
observations were recorded within the fiducial 30 minutes for CSS, two
times when only one reliable observation was made and two other times
when the usual four observations were made, but there was no reliable
detection. The magnitudes vary substantially indicating a transient of
some type to be determined.

CRTS also includes data from the Mt. Lemon Survey (MLS) which covers
mainly a narrow region of the sky near the ecliptic, and Siding Spring
Survey (SSS) which covers the Southern hemisphere. We have not
included data from MLS or SSS in the current study, but all methods
are equally applicable to them as well. About 75\% of the sky is
covered by CRTS, with parts near the poles and near the plane of our
Galaxy excluded.  Despite the relative sparsity of the CRTS
lightcurves, a strength of the survey is its longevity - we have data
where the epochs are spread over 10 years and hence there are parts of
the sky with several hundred observations making CRTS one of the
richest synoptic datasets. Our technique will scale to much larger
datasets including the 500 million lightcurves that CRTS now has. As
our method indicates which class an object belongs to based on its
lightcurve, we will be able to find matches for known classes. Our
measures are sufficiently informative to identify objects that do not
belong to known classes by using cluster analysis. This can in turn
lead to the discovery of newer classes as well as rarer counterparts
of known classes of astronomical objects.  CRTS images are available
for the entire dataset. When rare or unusual objects are found these
can be compared against the images to ensure that no artifacts or
spurious features have led to the object being an outlier. The whole
process can be automated and human intervention required only at
critical junctures like spectroscopic confirmation. Here we apply the
method to a large representative sample of transients together with
enough non-transients to make the comparisons useful.

Our strategies in deriving these critical measures are 1) selecting a
collection of relatively balanced, representative lightcurves of
various types and scales (\S2); 2) exploring the signatures
of these lightcurves (also in \S2); 3) developing a Gaussian process
regression model for the lightcurves with appropriate priors using
astronomical information and an empirical Bayesian approach (\S3); and
then 4) deriving new measures representing characteristics of the
lightcurves using the posterior mean regression curve and residuals.
These model-based measures complement existing measures. To examine
the power of the new measures in classifying lightcurves in
comparison to existing measures, five popular classification
procedures are used in four schemes of classification problem in \S4:
Linear Discriminant Analysis, Decision Trees, Support Vector Machines,
Neural Networks, and Random Forests. The results show that our
measures perform better than the existing measures. A discussion for
why our approach works better is given in \S5. Although our modeling
approach has been used for an astronomy application, the method could
be valuable for other applications involving the
classification of sparse, irregularly sampled time series with missing data.

\section{Data}
\label{sec:data}

We have selected a moderate sample of lightcurves with which to illustrate our
methods.  The measures available in each lightcurve are Right Acension (RA)
and Declination (Dec) which provide the position of the object on the sky,
Epoch as Julian Date, magnitude (negative logarithm of flux), an error estimate
on the magnitude.  The total number of lightcurves considered is 3720. The
selection is described below and summarized in Table~\ref{tab:nlc}.

We started with just the transients detected by CRTS in real-time over
about five years. These include {\em Active Galactic Nuclei} (AGNs),
{\em Blazars, Cataclysmic Variables} (CV), {\em Flare} stars and {\em
  Supernovae} (SNe), representing five very different types of
lightcurves (e.g. \citet{Djorgovski2011}).  We also included a set of
15 random pointings and objects within 3' of those pointings. These
objects are assumed to be non-transients because any transients in
there would have been detected earlier. The transients tend to be
fainter than typical objects (by definition - it is easier to catch
objects that are not normally seen but brighten and become visible for
a short duration). In order to offset that, two classes of brighter
variable objects were included --- Cataclysmic Variables from the {\em
  Downes} set (\citet{Downes2005}) and {\em RR Lyrae} which are
periodic variables with a period of $\sim$1 day. For the purposes of
this article, we have one class called \emph{non-transient} and seven
classes which we call \emph{transients} viz. AGN, Blazars, CV, Flares,
SNe, CV Downes, and RR Lyrae. Note that among the labelled types we
have considered here, only the \emph{RR Lyrae} are periodic. There are
methods for distinguishing periodic objects from non-periodic ones but
these are not addressed in this article.

Real-world data are much more assymetric and unbalanced than what we
have considered here. If we used a simple random sample, there would be very few
transients. Training on samples with enough
representatives for all classes would be an immense task. We include
sufficient non-variables to ensure that the dominant class
is represented but not excessively so. In CRTS, the latest catalogs are compared
with individual as well as combined catalogs from the archives. The
objects that have changed in brightness above a certain threshold
(well over a magnitude i.e. approximately a factor of 2.5) are marked
as transients. Only a few are found each night compared to millions of
nearly non-variable objects.

\begin{table}
\centering
\begin{tabular}{|ccccc|cc|c|} \hline
   \multicolumn{7}{|c|}{Functionally transient} & \\ \cline{1-7}
   \multicolumn{5}{|c|}{Transients}  & \multicolumn{2}{c|}{Bright variables} & \multicolumn{1}{c|}{}\\\cline{1-7}
   AGN & Blazar & CV  & Flare & SNe & CV Downes & RR-Lyrae & non-transient \\\hline
   140 & 124    & 461 & 66    & 536 & 376       & 292    & 1971 \\\hline
\end{tabular}
\caption{
Number of lightcurves for each transient type and non-transient objects
}
\label{tab:nlc}
\end{table}

The number of observations for each lightcurve varied greatly with a maximum of 641. The median length was 52. We excluded
lightcurves with fewer than five observations as these cannot reasonably be classified. The earliest date for any of these lightcurves was 53464 Julian Day (JD) and the latest was 56228, i.e. 5th April 2005 to 28th October 2012.
We have used 53464 as our zero-point and referred to all dates as number of days
beyond this.  Our set spans 2764 days.

\begin{figure}[htbp]
\begin{center}
\includegraphics[width=1.0\columnwidth]{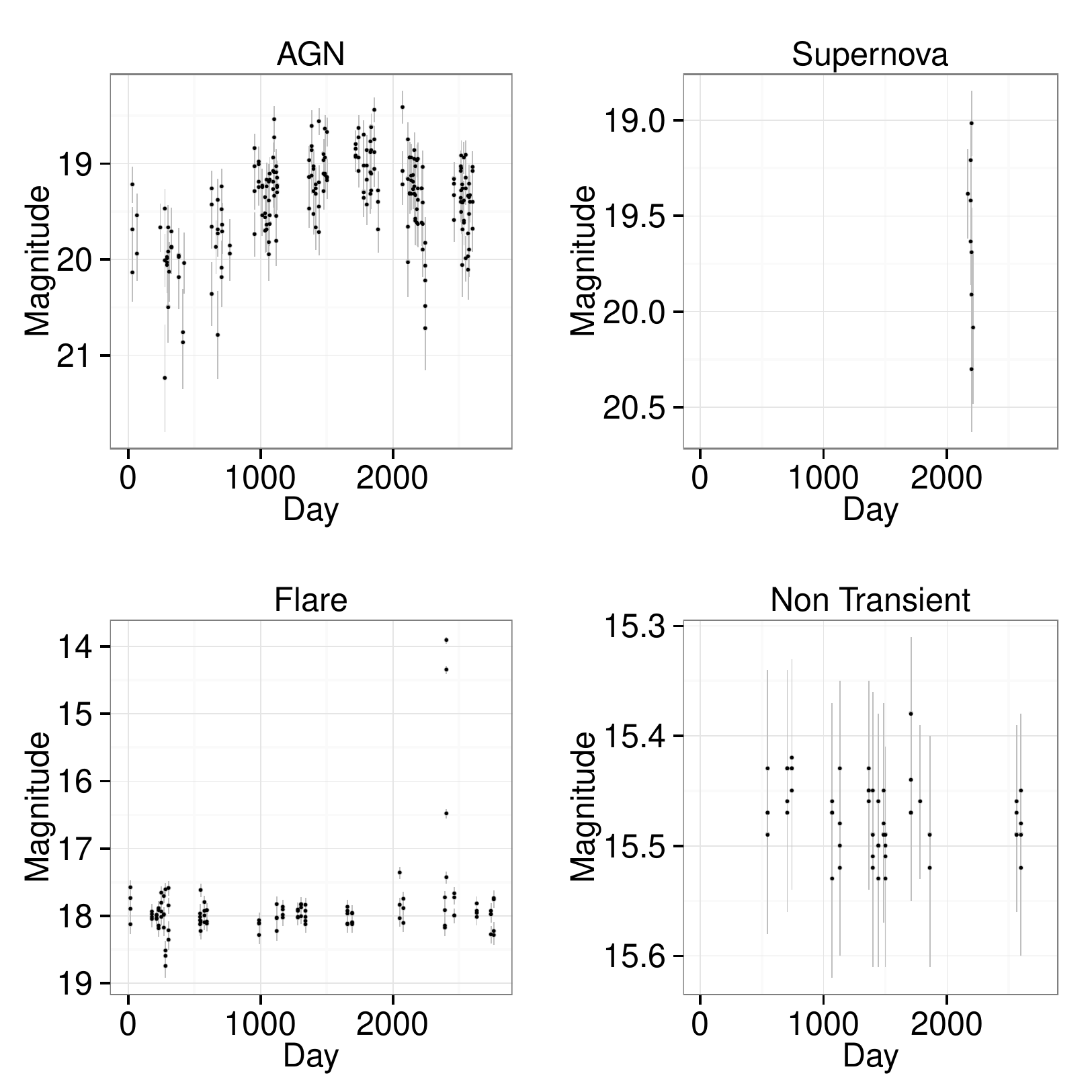}
\caption{Examples of four lightcurves: (i) CSS071216:110407-045134, an AGN (ii) CSS110405:141104+01115, a Supernova (iii) CSS111103:230309+40060, a Flare and (iv) 301904800767, a non-transient. Note that the range of the vertical axis varies.
}
\label{fig:lceg}
\end{center}
\end{figure}

An examination of the data is helpful in deciding which methods of analysis may
be appropriate.  In Figure~\ref{fig:lceg}, we see four examples of lightcurves. The objects are identified by their catalogue numbers for reference.

Understanding the pattern of measurement is crucial to proper modeling of these curves. The first example shows some gaps in an otherwise dense sequence of measurements. No observations were taken during these periods because the orbit of Earth precluded it. In the second example, there are no observations outside of a narrow range. Observations were attempted at other times but the object was too faint to be observed with a magnitude less than the survey astronomical detection limit of around 20.5. In the third example, we are fortunate that the spike in brightness was observed as this occurs during a brief period of time. In the fourth example, there are quite long periods with no observations but it seems reasonable to assume that no substantial variations in magnitude occurred during these periods given the nearly constant values of magnitude.

It would be useful to know exactly when observations were attempted for given objects while below detection limit. For the purposes of this analysis, we shall assume that all the objects may be surveyed throughout the period of the study but failures to observe have not been recorded. It will be clear how this information could be incorporated into our methods and that this would improve our results in \S3.

\section{Methods}
\label{sec:methods}

The nature of the data and the requirements of object classification
impose some constraints on what methods are practical. The
problem could be viewed as one of functional data analysis (see
\citet{ramsay:05}). However, there are several obstacles to pursuing
this approach. The observations on the lightcurves are very
irregular, both in time and in number. There are methods for dealing
with such data but there is a more serious obstacle in that there is
little sense in which the curves can be registered or
aligned. Excepting the rare case where objects are close in the sky
and measurements are likely to be correlated due to atmospheric conditions, lightcurves are
independent. This prevents us from using the ``borrowing
of strength'' that registration would allow.

This leads us to another style of analysis based on sample statistics.  Judgement is used to devise
statistics that measure various characteristics of the observed curves which we may
believe important in distinguishing them. We prefer that
these measures be relatively simple so that they can be applied
quickly and reliably for both short and long lightcurves.

About 20 measures are presented in \citet{Richards2011} that are mostly
derived from previous articles. They found these various measures to
be helpful in distinguishing objects. Since these measures have been
widely tested, at least for brighter data, we use these as a baseline
for our analysis. Our objective is to find additional measures that
improve the classification accuracy beyond this set.  For ease of
reference, we will call this set the \emph{Richards measures}. The
specific Richards measures we have used from Table 5 of \citet{Richards2011} are
moment-based measures: \texttt{skew}, \texttt{kurtosis}, \texttt{std}
and \texttt{beyond1std}, and magnitude-based measures:
\texttt{amplitude}, \texttt{maxslope}, \texttt{mad}, \texttt{medbuf},
\texttt{pairslope} and \texttt{rcorbor}, and percentile-based
measures: \texttt{fpr20}, \texttt{fpr35}, \texttt{fpr50},
\texttt{fpr80}, \texttt{peramp} and \texttt{pdfp}. We omitted the
linear trend measure as this was only large for lightcurves with few
observations so it becomes a substitute for a short curve measure. As
it happens, including it would not make much difference to the results
we present later. We coded these measures from the definitions in
\citet{Richards2011}.

Although the Richards measures encompass a wide variety of measures,
they do not use any concept of modeling the curves. The primary
innovation of this paper is to use such modeling to generate
additional measures. For lightcurve $i$, we posit a true underlying
curve $f_i(t)$ that we would see if we could observe the object
continuously without error. However, we are able to observe the object
only at times $t_{ij}$ for $j=1, \dots n_i$. Note that the times of
measurement may be almost the same for objects close in the sky but
quite different for objects which are farther apart.  We observe only
$y_{ij}$ for $j=1, \dots n_i$. We assume
\begin{equation}
  y_{ij} = f_i(t_{ij}) + \epsilon_{ij}
\end{equation}
where the errors $\epsilon_{ij}$ are normal with mean zero but will be correlated.

We considered several methods for estimating $f$ but found that
Gaussian process regression was the only satisfactory solution
compared to standard non-parametric methods like smoothing splines,
kernel smoothing or Lowess for the following reasons:
\begin{enumerate}
\item We have censored data - the lightcurve can fall below the
  detection limit during the range of observation. Standard methods do
  not deal with this. They can fit curves where we have data but they
  will not produce sensible fitted curves outside this range.
\item Sometimes we have only a handful of observations and but for
  other curves we may have hundreds. Simple parametric methods work
  well with small datasets while larger datasets require the
  flexibility of nonparametric methods. But Gaussian rocess
  regression works well for both.
\item We know the measurement error. This information is easily incorporated
  into the Gaussian process regression but there is no obvious way to take advantage
  of this information in the standard methods.
\end{enumerate}

\subsection{Gaussian Process Regression}

See \citet{rasmussen06:_gauss_proces_machin_learn} for a general introduction. This method requires
  that we specify a prior for the Gaussian process: $f(x) \sim GP(\psi(x),k(x,x^\prime))$. We use the popular squared
covariance kernel:
\begin{equation}
k(x,x^\prime) = \sigma^2_f \exp \left( - {1 \over 2l^2} (x-x^\prime)^2 \right) + \sigma^2_n \delta(x - x^\prime)
\end{equation}
where $\delta(x)$ is $0$ when $x \neq 0$ and $1$ when $x = 0$. Other reasonable choices of kernel are possible but we found the
results were similar (see online appendix for details).
One advantage of the Gaussian
process approach is that an explicit solution for the posterior is available that can be rapidly
calculated. We will need to classify large numbers
of future lightcurves as the measurements are collected so we need efficient methods.

\begin{figure}[htbp]
\begin{center}
\includegraphics[width=1.0\columnwidth]{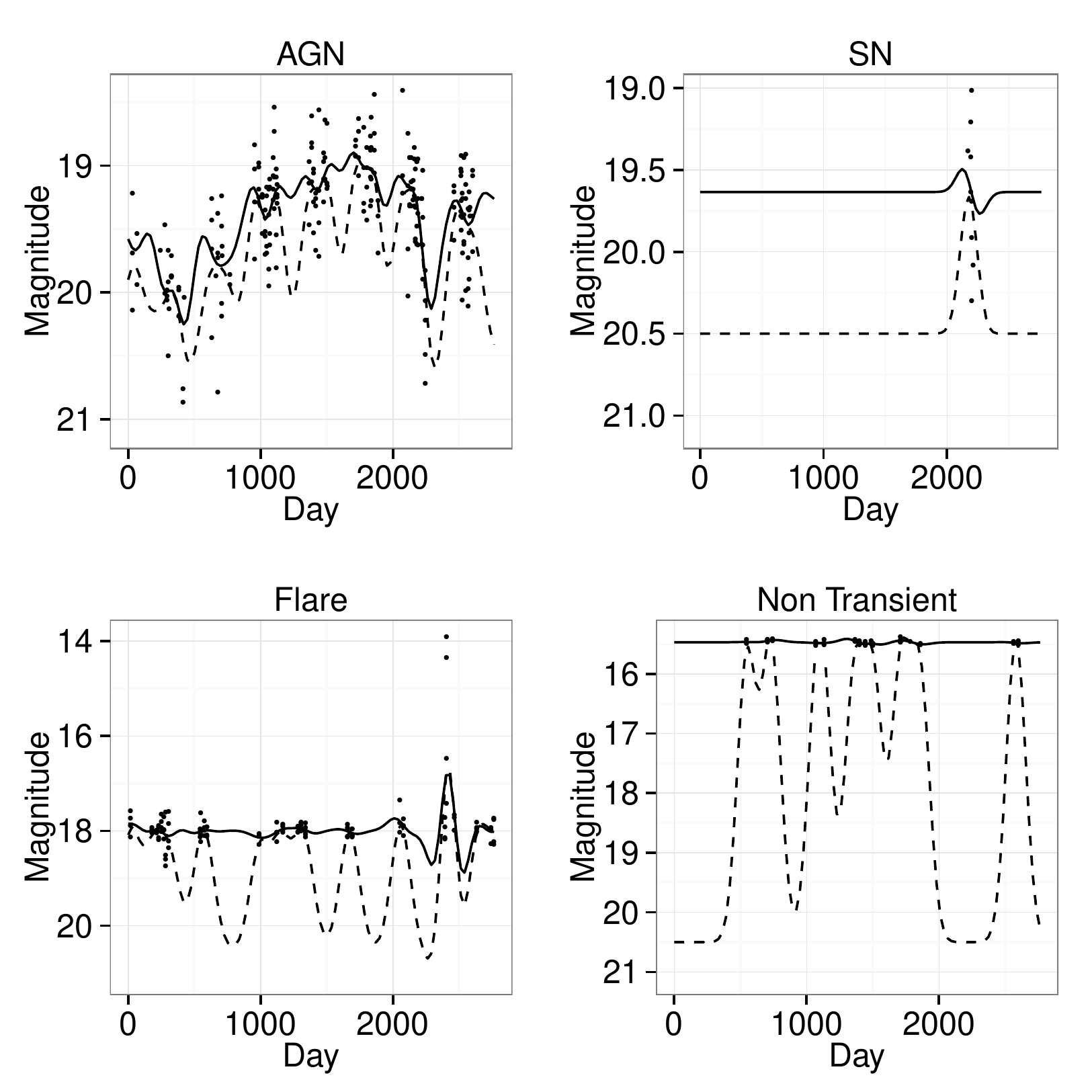}
\caption{Gaussian process regression fits to lightcurve data. The same four cases as in Figure~\ref{fig:lceg}. The solid line fit derives from a prior mean set at the median magnitude while the dashed line fit corresponds to a prior set at a magnitude of 20.5.}
\label{fig:gpr}
\end{center}
\end{figure}

There are four components of the prior which must be specified:

\begin{enumerate}
\item $\sigma^2_f$ is the signal variance. A very large fraction of objects to be classified in the future will be non-transients. These non-transients vary in signal but not very much. For this reason, we set $\sigma^2_f$ to be the
median observed variance in the non-transients.

\item $\sigma^2_n$ is the noise variance. Although it is uncommon in other applications, for astronomical data we are often able to estimate the measurement error. In this example, the measurement error varies a little from case to case. For simplicity, we
take the mean observed value of the measurement variance for $\sigma^2_n$.

\item $l$ is sometimes called the length-scale. It controls the amount
  of correlation and therefore the amount of smoothness in the
  resulting posterior fit. We use a value of $140$ days as seen
  in Figure~\ref{fig:gpr}. This choice is based on a subjective assessment on how much smoothness should be expected in these curves. Our classification performance is not very sensitive to this choice.

\item $\psi(x)$ is the prior mean. This choice is challenging and requires further discussion below. We take an empirical Bayes approach.
\end{enumerate}

We illustrate the issues in setting the prior mean in
Figure~\ref{fig:gpr}. What values are expected for
the curve in regions where there are no measurements? One answer is
that we might expect about the same magnitude as that seen elsewhere
for this object. This suggest setting the prior to the median
magnitude for the object. This choice can be seen in the solid line
fits in Figure~\ref{fig:gpr}. This works well enough in three cases
but fails for the supernova example because we do not expect this
curve to follow a similar magnitude at other times. If it did, it
would have been seen. So an alternative approach is to set the prior
to the detection limit at a magnitude of 20.5. This gives the dashed line fits as
seen in Figure~\ref{fig:gpr}. This works well for the supernova case but is
problematic for the other three curves. In regions of no measurement,
the fitted curve is drawn down towards the detection limit. We can
counteract this by increasing the length-scale (i.e. increasing the
smoothness in the prior) but tends to attenuate real effects and still
does not work well for relatively sparsely measured curves (such as
the non-variable in this example).

Our solution is to use an adaptive prior. When there is less than one
year of observations, we use the detection limit, otherwise we use
the median magnitude. The choice of a year is large enough that
sparse but widely measured curves such as the fourth example do not
use the detection limit. But the choice is small enough that the
detection limit is used in cases like the third example.  Now using
the data to select the prior may make our method at least partly
empirical Bayes rather than pure Bayes, but we need to judge the
method by its classification performance which is improved by this
choice.

\subsection{Curve Measures}

Given the posterior mean $\hat f$, we compute fitted curve measures from $\hat f_i$ for curve $i$ computed on an evenly spaced grid of values on the range of observation $u_{j}$ for $j=1, \dots, m=300$. The italicized word is the name of the variable for future reference:
\begin{itemize}
   \item \emph{totvar} total variation:  $\sum_j |\hat f_i(u_{i,j+1}) - \hat f_i(u_{ij})|/m$
   \item \emph{quadvar} quadratic variation: $\sum_j (\hat f_i(u_{i,j+1}) - \hat f_i(u_{ij}))^2/m$
   \item \emph{famp} amplitude of fitted function: $\max_t \hat f_i -  \min_t \hat f_i$.
   \item \emph{fslope} maximum derivative in the fitted curve: $\max_t |\hat f^\prime_i |$
\end{itemize}
We also use the maximum in absolute value of the scaled residuals from the fit, called \emph{outl}, as a measure.

Another feature of this data is the clustering of times of measurement
which can occur in groups of up to 4 observations that are spaced by
ten minutes within a thirty minute period. The Gaussian process
regression is not able to model the variation at this finer scale
because setting the length-scale $l$ to a much smaller value would
result in too rough a fit overall. We need another set of measures to
capture the characteristics at this scale of measurement.

We compute the mean within each of these groups of up to four observations as $\breve f_{ij}$ and
then compute the following measures:
\begin{itemize}
\item \emph{lsd}: the log of the standard error,  $\breve \sigma$, computed using the residuals from these group mean fits.

\item \emph{gtvar}: The group total variation $\sum_j |\breve f(t_{i,j+1}) - \breve f(t_{ij})|/n_i$

\item \emph{gscore} $\sum_j \phi ((\breve f_{ij} - \bar f_i)/\breve \sigma)/n_i$ where $\phi$ is the standard normal density, $\bar f$ is mean of the fitted group means.
\end{itemize}
The last measure is motivated by scoring methods used to judge prediction performance.

There are also some gaps within the Richards measures set of sample curve summary measures. We add the following:
\begin{itemize}
   \item \emph{shov} mean of absolute differences of successive observed values: \newline $\sum_j |y_{i,j+1} - y_{ij}|/n_i$
   \item \emph{maxdiff} the maximum difference of successive observed values: \newline $\max_j |y_{i,j+1} - y_{ij}|$
   \item \emph{dscore} the density score: $\sum_j \phi ((y_{ij} - \tilde f_i)/s_{ij}))/n_i$ where $\tilde f_i$ is the median observed magnitude for curve $i$ and $s_{ij}$ is the observed measurement error at $t_{ij}$.

\end{itemize}

There are other measures that may be informative for the current data we are
analyzing but may not have predictive value in future examples. We
have avoided using such measures.  They fall into three categories:

\begin{enumerate}
\item Measures based on the number of observations in a lightcurve. Some phenomena,
  such as supernovae, are not recurrent and subsequent observations
  may fall below the observable limit.  Lightcurves in such cases can be quite
  short but this is known only in retrospect so this is not usefully
  predictive. The number of observations does have some impact on the choice of prior and in scaling some of the measures, but we refrain from using this number (or anything closely related) as a direct measure for classification purposes.

\item The classification of an object should be invariant to the
  addition of a constant to the observed magnitudes. But some biases
  in the way that our example data was extracted would cause, say, the
  mean magnitude, to be an effective discriminator among the
  types. This mean magnitude will not be reproducible in future samples so we do not
  use this measure or anything related to it.

\item Location in the sky. The method of constructing our example
  dataset would mean that location would become useful
  discriminator. As it happens, location would provide some usable
  information for classification as extra-galactic objects are more
  likely to appear away from the galactic plane but we refrain
  from using this information here.
\end{enumerate}

There is additional information such as the nearest radio source or
the nearest galaxy which could also be useful in classification but we
do not use here.  We experimented with a larger set of additional
measures (as can be seen in the online appendix) but we have presented
only those that appear to have some additional value for
classification.

Given this set of measures, we can use any number of classification
methods to distinguish objects using lightcurves.  We demonstrate the use of our
measures using five popular classification methods. We will
generally use the default choice of options for the particular
implementation in $R$. Our objective is to show that our measures represent an
improvement over using the Richards measures alone. It is likely that the classification methods could be better tuned to
obtain a better result or that the reader may favor another
classification method. But that is not
the point of this article. We are not trying to claim one
classification method is better than another, just that our measures are better.

The methods we have used are:
\begin{description}
\item[LDA] Standard linear discriminant analysis method as
  implemented in  \citet{venables02:_moder_applied_statis_s}.
\item[TREE] Recursive partitioning as implemented in the
  \texttt{rpart} package  by \citet{therneau13}.
\item[SVM] Support vector machines as implemented in the
  \texttt{kernlab}
  package by \citet{karatzoglou04}.
\item[NN] Neural network as implemented in the \texttt{nnet} package
   by \citet{venables02:_moder_applied_statis_s}
\item[RF] Random forest ensemble implemented by the \texttt{randomForest} package by \citet{liaw02:_class_regres}
\end{description}

We log-transformed the measures that have extreme
skewness in order to improve classification performance. The same
transformations were used in all the comparisons below. Without these
transformations, both sets of measures would perform less well in
general for methods LDA, SVM and NN. The partitioning-based methods, TREE and RF,
are invariant to monotone transformations. Explicit details of the implementation
can be found in the Appendix.

\section{Results}
\label{sec:results}

Classification methods usually do not perform as well as expected when
applied to new data. When the same data are used to both fit and
evaluate a method, the classification rate is inflated. To avoid this
problem, we randomly split the data into 2/3 for training i.e. used to
develop the classification rule and 1/3 for testing, that is to
evaluate how well the rule performs. Since we are only interested in
the relative performance of the classification measures and methods
and because the sample size is relatively large, we present only one
random split. In the online appendix, we repeat the calculations
for 100 random splits and the results are not qualitatively different.

\subsection{Classification Performance}
\label{sec:class-perf}

We considered four different types of classification problem with bold labels used for future reference:
\begin{description}
\item[All] The overall problem of classifying eight types --- the
  non-transients and the seven transient types.

\item[Transient or not] Perhaps the first step in any lightcurve
  classification process will be to determine which objects are transient.

\item[Transient only] Having separated out the non-transients, the next step
  might be to identify the type of the transient. For this problem, we
  delete the non-transients from both the test and training data.

\item[Heirarchical] An alternative approach to classifying all objects
  directly is to first classify objects into transient or not
  transient, then if transient to classify among the seven available
  types.
\end{description}

We show the percentages correctly classified using the Richards
measures in Table~\ref{tab:rich} and using our measures (which
incorporate the Richards set) in Table~\ref{tab:our}.

\begin{table}
\centering
\begin{tabular}{rrrrrr}
  \hline
 & LDA & TREE & SVM & NN & RF \\
  \hline
All & 56.7 & 58.6 & 66.1 & 63.3 & 67.3 \\
  Transient or not & 74.7 & 79.5 & 81.0 & 75.2 & 82.5 \\
  Transient only & 54.5 & 58.9 & 64.4 & 60.1 & 62.9 \\
  Heirarchical & 56.4 & 60.4 & 64.7 & 58.8 & 65.6 \\
\hline
\end{tabular}
\caption{Percentage correctly classified using the Richards measures.
\label{tab:rich}}
\end{table}

\begin{table}
\centering
\begin{tabular}{rrrrrr}
  \hline
 & LDA & TREE & SVM & NN & RF \\
  \hline
All & 76.0 & 71.9 & 80.2 & 79.6 & 80.5 \\
  Transient or not & 90.4 & 88.4 & 92.0 & 91.6 & 91.8 \\
  Transient only & 70.1 & 65.1 & 74.3 & 72.3 & 74.2 \\
  Heirarchical & 76.0 & 72.7 & 79.9 & 78.5 & 79.8 \\
   \hline

\end{tabular}
\caption{Percentage correctly classified using our measures in addition to the Richards set.
\label{tab:our}}
\end{table}

The standard error for the classification rate is just less than 1\%
which is helpful in judging which differences are notable in these
tables.  Table~\ref{tab:rich} and Table~\ref{tab:our} show that our
measures provide a significant improvement to the Richards measures
alone which might be regarded as the previous state of the art. Of
course, adding additional measures can only improve the fit of a
model, but we are using an independent test set so we can be sure the
improvement is more than illusory.  There is little to distinguish the
heirarchical approach from the one-step method although we would
recommend the heirarchical approach on an unbiased sample of
lightcurves because these would be dominated by non-variables.

Tables~\ref{tab:confusionrich} and \ref{tab:confusionour} show the numbers of objects classified
into all eight types compared to their actual types. We present only the random forest results
as this was the best performing method. The most noticeable differences between the two sets of
measures is that our method results in classifications in all eight types while the older
set fails to classify any objects into four of the types. Since the Downes set is just another form
of CV, we not so concerned about a failure to distinguish these two. We can see that
Flares are hard to identify.

\begin{table}
  \centering
  \begin{tabular}{r|rrrrrrrr}
 & \multicolumn{8}{c}{Actual types} \\
Predicted  &  AGN &  Blazar &  CV &  Downes &  Flare &  NT &  RR-Lyrae &  SNe  \\ \hline
AGN &  0 &  0 &  0 &  0 &  0 &  0 &  0 &  0     \\
Blazar &  0 &  0 &  0 &  0 &  0 &  0 &  0 &  0          \\
CV &  5 &  26 &  95 &  53 &  4 &  20 &  3 &  27        \\
CV Downes &  0 &  0 &  0 &  0 &  0 &  0 &  0 &  0         \\
Flare &  0 &  0 &  0 &  0 &  0 &  0 &  0 &  0                                \\
non-transient (NT) &  31 &  7 &  26 &  73 &  16 &  497 &  47 &  80          \\
RR-Lyrae &  0 &  1 &  2 &  3 &  0 &  12 &  53 &  3                              \\
SNe &  8 &  5 &  22 &  6 &  4 &  29 &  0 &  82                                \\
  \end{tabular}
  \caption{Confusion matrix for the Richards set using random forest classification of all eight types. The rows are the predicted types while the columns are the actual types.
    \label{tab:confusionrich}}
\end{table}

\begin{table}
  \centering
  \begin{tabular}{r|rrrrrrrr}
 & \multicolumn{8}{c}{Actual types} \\
Predicted  &  AGN &  Blazar &  CV &  Downes &  Flare &  NT &  RR-Lyrae &  SNe  \\ \hline
AGN                     &  31 &  3 &  0 &  2 &  0 &  2 &  0 &  2                   \\
Blazar                  &  0 &  27 &  3 &  7 &  0 &  0 &  0 &  0                   \\
CV                 &  2 &  4 &  93 &  26 &  0 &  4 &  2 &  14                 \\
CV Downes                   &  1 &  2 &  15 &  58 &  0 &  7 &  5 &  0                  \\
Flare             &  0 &  0 &  0 &  3 &  8 &  0 &  0 &  0                    \\
non-transient (NT)          &  8 &  0 &  9 &  25 &  15 &  541 &  1 &  16               \\
RR-Lyrae                  &  0 &  1 &  0 &  7 &  0 &  0 &  95 &  0                   \\
SNe                    &  2 &  2 &  25 &  7 &  1 &  4 &  0 &  160                 \\
  \end{tabular}
  \caption{Confusion matrix for our measures using random forest classification of all eight types. The rows are the predicted types while the columns are the actual types.
  \label{tab:confusionour}}
\end{table}

Our sample heavily over-represents transients which would constitute
less than 1\% of an unbiased sample. Hence, even a null method which
classified randomly based on prior proportions would achieve around
99\% accuracy. Certainly any sensible method will do even better than
this and it would take a very large unbiased sample to distinguish
different methods. This explains why we have used a more balanced,
although biased, representation of the eight types. Similar strategies
are used in case-control studies.

Because these classification methods will be applied to very large
numbers of objects, even quite low error rates will result in large
numbers of misclassified objects resulting in wasted resources or
missed opportunities. For this reason, the primary classification into
transient against non-transient is particularly important. We can see
our proposed measures perform well in this respect, halving the previous error
rate, although there remains further room for improvement.

\subsection{Testing on Fresh Data}
\label{sec:testing-fresh-data}

We verified the performance of our measures by applying the
methodology to two datasets distinct from the training set. We
assembled 574 transients that have been more recently manually
classified consisting of the types AGN, Blazar, CV, Flare and SNe. This
is a complete set of CSS transients from 2013 where astronomers using
additional auxiliary data were reasonably confident of the nature of
the transient. We used all the objects of these five types from the
original set of data (updating them to include more recently collected
observations to match the lightcurve durations) to train classification
rules. We then applied these rules to a test set formed from the
recent set of transients. The classification rates from this exercise a
shown in Table~\ref{tab:fresh5}. The addition of our set of measures results in an
improved performance over the Richards set alone.
\begin{table}
\centering
\begin{tabular}{rrrrrr}
  \hline
 & LDA & TREE & SVM & NN & RF \\
  \hline
Richards & 60.8 & 64.0 & 72.6 & 71.1 & 73.1 \\
Ours+Richards & 74.6 & 67.4 & 78.0 & 77.2 & 79.0 \\
\hline
\end{tabular}
\caption{Percentage of the new set of five transient types correctly classified.
\label{tab:fresh5}}
\end{table}

We also assembled a large set of 100,000 lightcurves from CSS. We sampled 50,000 objects from $0 < S_J < 0.01$ and labeled these non-variables, and 50,000 more from $0.5 < S_J < 1$ which we labeled as variable. Any labeling based on a single variable will not be entirely reliable so we have chosen ranges from the extremes of $S_J$ to increase confidence that the labels are correct. See \citet{Drake2014} for background.
 Very few of the
variables are likely to be transients and most of them will be of types not
already considered earlier. We randomly divided the data into two equal
parts. One half was used to train classification rules and the other half to
test the performance. The classification rates are shown in Table~\ref{tab:big} where we see
that the addition of our set of measures to the Richards set greatly improves
the rates.
\begin{table}
\centering
\begin{tabular}{rrrrrr}
  \hline
 & LDA & TREE & SVM & NN & RF \\
  \hline
Richards & 75.5 & 79.4 & 85.3 & 76.8 & 85.5 \\
Ours+Richards & 97.5 & 89.2 & 98.4 & 98.3 & 96.8 \\
\hline
\end{tabular}
\caption{Percentage of 50,000 variables and non-variables correctly classified.
\label{tab:big}}
\end{table}

\subsection{Feature Selection}

\begin{figure}[htbp]
  \centering
  \includegraphics[width=1.0\columnwidth]{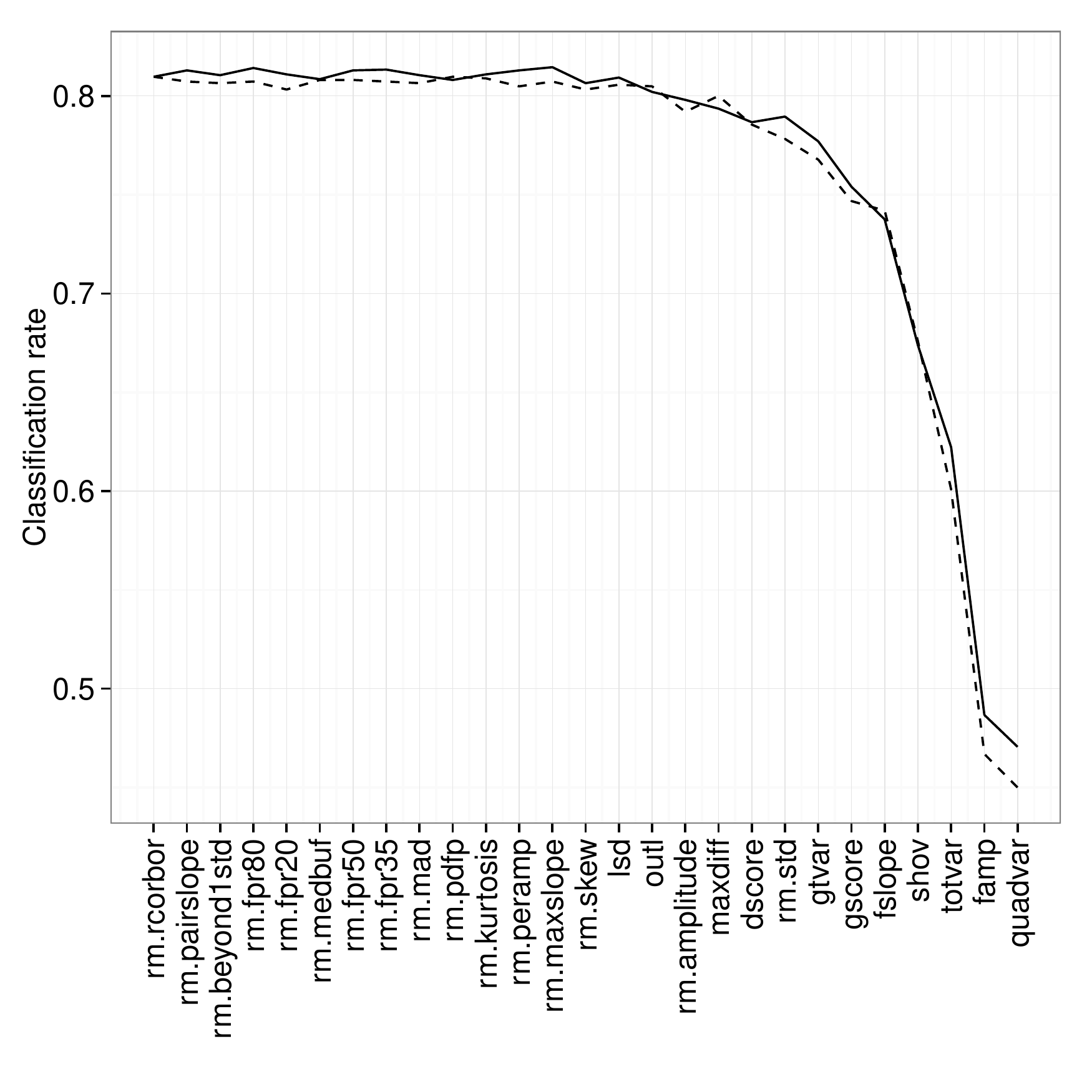}
  \caption{Stepwise selection of measures. The classification rates
    (solid for the training set and dashed for test set) are shown
    after the named measure is removed from the model. The order in
    which the measures are removed is determined by the least decrease
    in the Gini index. Measures from the Richards set have the prefix
    \emph{rm}.}
  \label{fig:vs}
\end{figure}

The random forest method provides a means of determining the worth of
predictors. Suppose that within a particular node, the proportion
classified as type $i$ is given by $p_i$. The Gini index is defined
as $1-\sum_i p_i^2$ and can be used as a measure of node impurity. We can see
how much the Gini index averaged across nodes decreases when a
measure is removed from the current set. We remove the measure which
leads to the least decrease. We refit the model with the reduced set
and repeat the process until all measures have gone.  The classification rate
after each measure is removed is shown in Figure~\ref{fig:vs} for the problem of
classifying all 8 types.  See also \citet{Donalek2013} for similar
procedures.

There is little difference in classification accuracy between the
training and test datasets which is a good indication that we are not
over-fitting.  We see that this selection process removes most of the
older non-model based measures without any noticeable loss in
classification accuracy. Our fitted curve measures \emph{quadvar}, \emph{famp},
\emph{totvar} and \emph{fslope} are among the most useful classification
variables. Hence we can see that deriving measures based on our
model is a good place to start and not merely a way to supplement
existing older measures. The message from the other three
classification problems is similar. We do not propose that
the older measures be discarded as they are generally simple to
compute and may be useful in new situations.

\section{Discussion}
\label{sec:conclusion}

We have presented two advances. In Statistics, we have shown how
Gaussian process regression can be adapted and the corresponding
priors developed to deal with data of varying sampling density,
structures and scales. We are able to deconvolute the underlying
curves $f_i$ and measurement noise $\epsilon_i$ to distinguish the objects rather than
use summary statistics that use samples $y_{ij}$ with mixed up noise and
signal. With some further effort we might show that the measures based
on our estimated curves, under some regularity conditions, would be
consistent for the features of the true underlying curves. A summary
statistic-based approach can be biased or inconsistent for these
measures.  In Astronomy, we have demonstrated a new method of
generating measures representing features of lightcurves that are
significantly better in classifying objects than previous methods.

There is further scope for improvement in performance by optimising
the classification using routine methods. With more detailed
information about when locations were surveyed but no object observed,
we can further refine our priors to obtain superior results.  The
measures we have developed are now being used for several purposes. We
can apply the method for new data where the location has only been
surveyed for a shorter period of time. The measures can all be scaled
appropriately. We have experimented by taking time-wise subsets of
this data and have found that although the absolute performance drops
with shorter curves, the relative performance over the older set of
measures remains. Furthermore, the measures provide the means to
detect objects of unknown type. By adding a richer and more powerful
set of measures, we have increased the potential for such interesting
discoveries. Some of our measures have already been used in classifying
new lightcurves.

\section*{Acknowledgements}
\label{sec:acknowledgements}

This work is one of results from the Imaging Working Group at the {\em SAMSI's 2012-13 Program on Statistical and Computational Methodology for Massive Datasets}. We thank SAMSI for bringing us together and for their financial support. The CSS survey is funded by the National Aeronautics and Space Administration under Grant No. NNG05GF22G issued through the Science Mission Directorate Near-Earth Objects Observations Program.  The CRTS survey is supported by the U.S.~National Science Foundation under grants AST-0909182 and AST-1313422. We are also thankful to the Keck Instutute of Space Studies, and part of the work was supported through the Classification grant, IIS-1118041. We thank SG Djorgovski for useful comments and AJ Drake and MJ Graham for help in assembling the 100K dataset.

\section*{Appendix}
\label{sec:app}

Our data, code and detailed results are available as a supplement to be found
at \newline \texttt{people.bath.ac.uk/jjf23/modlc}

\bibliography{modlc}

\end{document}